\documentclass[pra,twocolumn,showkeys,showpacs,10pt,superscriptaddress,aps]{revtex4-2}

\usepackage[english]{babel}
\usepackage{amsmath}
\usepackage{dcolumn}
\usepackage{amssymb}
\usepackage{textcomp}
\usepackage[utf8]{inputenc}
\usepackage[normal]{subfigure}
\usepackage{lipsum}
\usepackage{hyperref}
\usepackage[justification=raggedright]{caption}
\usepackage{tikz}
\usepackage{nicefrac}
\usepackage{siunitx}
\usepackage{physics}
\usepackage{float}
\usepackage{import}

\renewcommand{\vec}{\mathbf}

\DeclareMathOperator{\ee}{e}

\begin{document}
\title{Forming complex neurons by four-wave mixing in a Bose-Einstein condensate}

\newcommand{\affT}{Technische Universität Darmstadt, Institut für Angewandte Physik, Hochschulstraße 4a, 64289 Darmstadt, Germany}

\author{Kai Niklas Hansmann}
\email{kai.hansmann@physik.tu-darmstadt.de}

\author{Reinhold Walser}
\affiliation{\affT}

\date{\today}


\begin{abstract}
A physical artificial complex-valued neuron is formed by four-wave mixing in a homogeneous three-dimensional Bose-Einstein condensate. Bragg beamsplitter pulses prepare superpositions of three plane-waves states as an input- and the fourth wave as an output signal. The nonlinear dynamics of the non-degenerate four-wave mixing process leads to Josephson-like oscillations within the closed four-dimensional subspace and defines the activation function of a neuron. Due to the high number of symmetries, closed form solutions can be found by quadrature and agree with numerical simulation.
The ideal behaviour of an isolated four-wave mixing setup is compared to a situation with additional population of rogue states. We observe a robust persistence of the main oscillation. As an application for neural learning of this physical system, we train it on the XOR problem. After $100$ training epochs, the neuron responds to input data correctly at the  $10^{-5}$ error level. 
\end{abstract}

\maketitle

\section{Introduction}
Neural networks and deep learning methods have evolved dynamically into a far-reaching research field \cite{LeCun2015}. Nowadays, applications can be found in diverse areas like bio-chemistry \cite{Senior2020}, medicine \cite{Elmarakeby2021}, image analysis \cite{Ravindran2022}, computer-games \cite{Silver2016,Wurman2022}, gravitational-wave detection \cite{Heimann2023} and sundry more. There exists a variety of implementations of artificial neural networks: electronic implementations using graphical processing units \cite{Mittal2019,Mittal2019a}, but also other physical implementations can be considered \cite{Kaspar2021,Nautrup2022}. In particular, optical implementations receive a lot of attention \cite{Sui2020,Kuratomi1993,Gao1997,Psaltis1990,Cheng2019,Tait2017,Shen2017,Feldmann2019,Zhang2021,Ashtiani2022}.

The key issue in setting up a novel physical implementations of artificial neural networks is the description of their constituents, the artificial neurons. Diverse approaches can realize artificial neurons in photonic systems \cite{Zuo2019,Ryou2021,Miscuglio2018,Skinner1994,Dejonckheere2014}. 
In this paper, we consider an artificial neuron using the inherent nonlinearity of ultracold coherent bosonic matter-waves.

Coherent matter-waves show a wide range of nonlinear effects, which, for example, have been used to detect the phase transition towards a Bose-Einstein condensate (BEC) experimentally \cite{Anderson1995,Andrews1997}. For our purposes, we investigate the process of four-wave mixing (FWM) in coherent matter-waves, which is well-known from nonlinear optics \cite{Maker1965}. If phase-matching conditions are fulfilled in a nonlinear optical medium, three frequencies interact in a way such that an initially absent fourth frequency can be observed. Following the advent of the BEC, theoretical investigations \cite{Trippenbach1998,Trippenbach2000,Wu2000,SUN2004,Rowen2005}, as well as experiments \cite{Deng1999,Vogels2002} demonstrated the equivalent FWM process. There, momentum components of the BEC took over the role of optical frequencies from the initial scenario. In an idealized homogeneous BEC, we can show that the FWM process of plane waves exhibits Josephson-like oscillations \cite{leggett91,leggett01,Smerzi1997a,Williams1999,Kronjager2005,walser08,Grupp2013}. 

We utilize this highly nonlinear process to implement a complex-valued neuron, where we identify three momentum components as input and the fourth component as output. The input-output-relations of the neuron are highly nonlinear and will be investigated in detail. As an application, we train the FWM neuron on the benchmark XOR problem.

The paper is organized as follows. In Sec.~\ref{sec:ideal_4WM}, we introduce the isolated FWM problem in a three-dimensional homogeneous BEC, revealing the dynamics of populations and phases of the four momentum components. In 
Sec.~\ref{sec:josephson}, we solve the FWM dynamics analytically in form of Josephson oscillations. After the investigation of FWM under ideal conditions, we look at the influence of additional population in momentum components outside of the FWM manifold in Sec.~\ref{sec:background}.  Finally, we introduce the artificial FWM neuron in Sec.~\ref{sec:4WM_neuron}, discuss the nonlinear activation function of the neuron and introduce the steepest descent learning method for complex-valued neurons. As an application, we train the  FWM neuron on the XOR problem. In the appendix, we discuss the preparation of  FWM input states using Bragg pulses well known in atom interferometry \cite{Neumann2021,Ertmer2009,Marksteiner1995}.

\section{Ideal Four-wave mixing} \label{sec:ideal_4WM}

The dynamics of a weakly interacting BEC described by the order parameter $\psi(\vec{r},t)$ are given by the Gross-Pitaevskii equation \cite{Gross1961,Pitaevskii1961}
    \begin{align}
        i\hbar\partial_t\psi=&\left[-\frac{\hbar^2}{2m}\nabla^2+U+gn\right]\psi,\label{eq:GPE}
    \end{align}
where $n(\vec{r})=\vert\psi(\vec{r})\vert^2$ is the density,  $N=\int \text{d}^3r\, n$ is the total particle number, $U(\vec{r})$ describes an external potential and the  coupling constant  $g=4\pi\hbar^2a_s/m$ is proportional to the atomic s-wave scattering length $a_s$ and the mass of the atoms $m$. The Gross-Pitaevskii Lagrangian functional for such a system reads \cite{Hill1951,CohenTannoudji1997}
    \begin{align}
    \label{eq:schroedingerlagrangian}
        L=&\int \dd^3r\;(i\hbar\psi^*\partial_t\psi-\mathcal{E}), \\
        \mathcal{E}=&\frac{\hbar^2}{2m}|\nabla\psi|^2+Un+\frac{g}{2}n^2.
    \end{align}
    
In the following, we consider the case of a homogeneous BEC with $U=0$ and periodic boundary conditions. Then, a wave function $\ket{\psi}=\ket{\psi_\alpha}+\ket{\psi_\beta}$, is a coherent superposition of plane waves  $\ket{\vec{k}_j}$ with complex amplitudes $\alpha_j$ and $\beta_l$. It consists of a FWM state 
    \begin{align}
    \label{eq:FWMstate}
        \ket{\psi_\alpha}=
        \sum_{j=1}^4 \sqrt{N}  \alpha_j\ket{\vec{k}_j}
    \end{align}
and a residual wave 
    \begin{align}
        \ket{\psi_\beta}&=\sum_{l>4} \sqrt{N}\beta_l 
        \ket{\vec{k}_l},
    \end{align}
which is orthogonal $\braket{\psi_\alpha}{\psi_\beta}=0$ to the FWM state.

The complex amplitudes $\alpha_j$, in terms of absolute value and phase, are given by
	\begin{align}
		\alpha_j=\sqrt{n_j}e^{-i\varphi_j}. \label{eq:polar_decomposition}
	\end{align}
Thus, $n_j=\vert\alpha_j\vert^2$ is the probability to be in the momentum state $\ket{\vec{k}_j}$. The mode functions
    \begin{math}
        \braket{\vec{r}}{\vec{k}_j}=\ee^{i \vec{k}_j\vec{r}}/\sqrt{V}
    \end{math}
are normalized in a cuboid with lengths $(L_1, L_2, L_3)$ and a volume $V=L_1L_2L_3$. For periodic boundary conditions, the wave-numbers $k_j=2\pi \kappa_j/L_j$ are quantized with $\kappa_j\in\mathbb{Z}$ and the plane wave states are orthonormal $\braket{\vec{k}_i}{\vec{k}_j}=\delta_{ij}$. 

The conditions for FWM are energy and momentum conservation \cite{Deng1999}
    \begin{align}
        \omega_1+\omega_2&=
        \omega_3+\omega_4, &
        \vec{k}_1+\vec{k}_2&=\vec{k}_3+\vec{k}_4,
         \label{eq:4WM_cond}
    \end{align}
for the dispersion relation of massive particles
\begin{align}
    \omega_j=\omega(\vec{k}_j)=\frac{\hbar |\vec{k}_j|^2}{2m}.
\end{align}
Experimentally, momentum states fulfilling these conditions can be prepared using atomic beamsplitters based on Bragg diffraction \cite{Martin1988,Neumann2021}, as discussed in App.~\ref{sec:state_preparation}.

\begin{figure}[b]
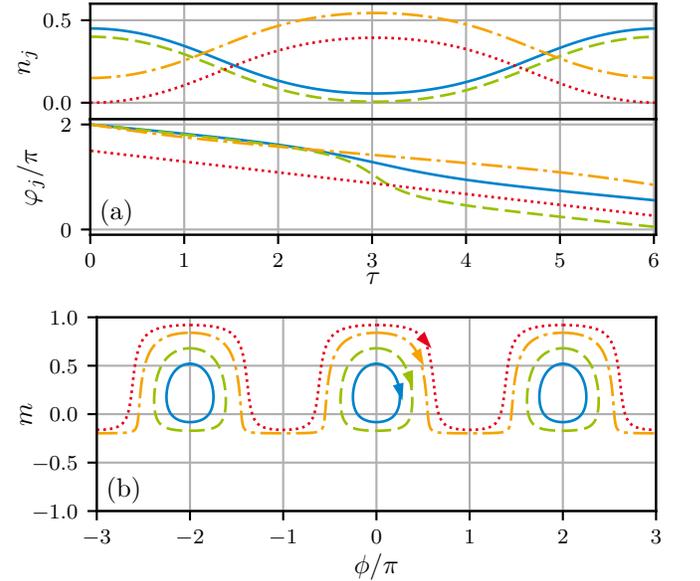

    \begin{subfigure}
        \centering
        \scalebox{1.0}{\import{Pictures/}{rk4_ode_1.pgf}}
    \end{subfigure}
    \begin{subfigure}
         \centering
        \scalebox{1.0}{\import{Pictures/}{rk4_ode_3.pgf}}
    \end{subfigure}
 	\caption{(a) Populations $n_j$ and phases $\varphi_j$ of the FWM states $\ket{\vec{k}_1}$ (blue, solid), $\ket{\vec{k}_2}$ (green, dashed), $\ket{\vec{k}_3}$ (orange, dash-dotted) and $\ket{\vec{k}_4}$ (red, dotted) versus dimensionless time $\tau$ for $\bar{\omega}_j=1$. 
 	(b) FWM trajectories $(\phi(\tau),m(\tau))$ in phase-space for the initial conditions  $\phi(0)=0$, $m(0)=0.52$ (blue, solid), $m(0)=0.68$ (green, dashed), $m(0)=0.84$ (yellow, dash-dotted) and $m(0)=0.92$ (red, dotted). The other constants of motion read	$m_{12}=n_1-n_2=0.4$ and  $m_{34}=n_3-n_4=0.02$.}
 	\label{fig:rk4_ode}
\end{figure}

In the ideal FWM scenario, the residual wave is absent, $\beta_l=0$. Consequently, $\sum_{j=1}^4n_j=1$. In this case, the nondimensionalization of the physical Lagrangian functional \eqref{eq:schroedingerlagrangian} is achieved by measuring time $\tau=\gamma t$ by a clock that ticks with frequency $\gamma=gN/\hbar V$, by scaling the frequencies $\bar{\omega}_j=\omega_j/\gamma$, as well as scaling and shifting the Lagrangian function $\mathcal{L}=1+VL/gN^2$. Thus, the mathematical Lagrangian functional reads
\begin{align}
\label{eq:lagarangenondim}
    \mathcal{L}&=\sum_{j=1}^4 i \alpha_j^\ast \dot{\alpha}_j-\mathcal{E}, \\
    \label{eq:energydiscrete}
         \mathcal{E}&=\sum_{j=1}^4\varepsilon_j+
    2(\alpha_1^\ast\alpha_2^\ast\alpha_3\alpha_4+
    \text{c.c.}),
\end{align}
where $\dot{\alpha}_j$ denotes $\partial_{\tau}\alpha_j$ and the mean-field shifted single particle energies $\varepsilon_j$ and chemical potentials $\mu_j$ are defined as
\begin{align}
    \varepsilon_j&=\bar{\omega}_jn_j-\frac{n_j^2}{2},
    &
    \mu_j&=\frac{\partial\varepsilon_j}{\partial n_j}=
    \bar{\omega}_j-n_j.
\end{align}

According to the Euler-Lagrange equations \cite{Kronjager2005}
\begin{align}
\dfrac{\text{d}}{\text{d}\tau}
\dfrac{\partial\mathcal{L}}{\partial{\dot{\alpha}}_j}
=\dfrac{\partial\mathcal{L}}{\partial \alpha_j},
\end{align}
the complex amplitudes evolve as 
\begin{align}
\begin{aligned}
    i\dot\alpha_1=\mu_1 \alpha_1+2\alpha_2^{\ast}\alpha_3\alpha_4,\\
    i\dot\alpha_2=\mu_2 \alpha_2+2\alpha_1^{\ast}\alpha_3\alpha_4,\\
    i\dot\alpha_3=\mu_3 \alpha_3+2\alpha_4^{\ast}\alpha_1\alpha_2,\\
    i\dot\alpha_4=\mu_4 \alpha_4+2\alpha_3^{\ast}\alpha_1\alpha_2.
\end{aligned}    
\label{eq:complexdgl}
\end{align}

Clearly, these equations are highly symmetric, which can be explored using the polar decomposition of the complex amplitudes (\ref{eq:polar_decomposition}). Most notably is the interaction term in \eqref{eq:energydiscrete}. It will coherently couple the subspaces 
$\{\ket{\vec{k}_1},\ket{\vec{k}_2} \}\longleftrightarrow \{\ket{\vec{k}_3},\ket{\vec{k}_4} \}$ through the relative phase-difference $\phi=\varphi_1+\varphi_2-\varphi_3-\varphi_4$ and population imbalance  $m=n_1+n_2-n_3-n_4$. To confirm this first impression, we evaluate the dynamics of the system \eqref{eq:complexdgl} numerically. 
In Fig.~\ref{fig:rk4_ode} (a), we show periodic oscillations of the populations $n_j(\tau)$ and quasi-linear evolution of the phases $\varphi_j(\tau)$. 
In the $\phi$-$m$ phase-space, one finds periodic as well as aperiodic orbits of a mathematical pendulum with a separatrix in between (see Fig. \ref{fig:rk4_ode} (b)). 
The population imbalance $m$ is constrained by the conserved quantities $m_{12}=n_1-n_2$ and $m_{34}=n_3-n_4$ discussed in Sec.~\ref{sec:josephson}.

\section{Josephson oscillations of four-wave mixing amplitudes} \label{sec:josephson}

\subsection{Coordinate transformation}

Due to the Lagrangian field theory, the time-independent Hamiltonian energy  
\eqref{eq:schroedingerlagrangian} and the FWM state ansatz \eqref{eq:FWMstate}, we obtain a discrete nonlinear set of four Hamiltonian equations with a number of symmetries. This constrains the dynamics to a two-dimensional phase-space, analogous to the mathematical pendulum. Due to the phase-invariant structure of the self-energy $ gn^2$, typical Josephson oscillations \cite{leggett91,leggett01,Smerzi1997a,Williams1999,Kronjager2005,walser08,Grupp2013} emerge. Similar equations appear in the study of semicassical methods in the theory of Rydberg atoms \cite{Braun1993}.
 
Guided by this idea, we introduce adapted coordinates
\begin{equation}
\begin{aligned}
    \alpha_1&=\sqrt{n_1}\ee^{-i(\Phi+\phi/4+\varphi)},
    &
    \alpha_2&=\sqrt{n_2}\ee^{-i(\Phi+\phi/4-\varphi)},\\
    \alpha_3&=\sqrt{n_3}\ee^{-i(\Phi-\phi/4+\theta)},
    &
    \alpha_4&=\sqrt{n_4}\ee^{-i(\Phi-\phi/4-\theta)}.
\end{aligned}    
\end{equation}
From the global phase invariance of \eqref{eq:schroedingerlagrangian} or \eqref{eq:lagarangenondim}, one finds that the total occupation $\sum_{j=1}^4 n_j=\text{const.}$ This can be used to construct a generating function $R(\alpha_1, \alpha_2, \alpha_3, \alpha_4, \Phi,\phi,\varphi,\theta)$ as
\begin{align}
R=\frac{i}{2}\ee^{2 i\Phi} \Bigl(&
\alpha_1^2\ee^{2i(\phi/4+\varphi)}+
\alpha_2^2\ee^{2i (\phi/4-\varphi)}
\nonumber\\
 &+\alpha_3^2\ee^{2i(-\phi/4+\theta)}+
\alpha_4^2\ee^{2i (-\phi/4-\theta)}\Bigr).
\end{align}
According to the rules of Hamiltonian mechanics \cite{Reineker2021}, this generating function relates old coordinates  $(\alpha_1, \alpha_2, \alpha_3, \alpha_4)$ to new coordinates $(\Phi,\phi,\varphi,\theta)$. In turn, one can obtain the old momenta
\begin{align}
\pi_j=\frac{\partial R}{\partial\alpha_j}=i \alpha_j^\ast,
\end{align}
as well as the new momenta
\begin{align}
    P_{\Phi}=&-\frac{\partial R}{\partial \Phi}=n_1+n_2+n_3+n_4,\\
    P_{\phi}=&-\frac{\partial R}{\partial \phi}=\frac{n_1+n_2-n_3-n_4}{4}\equiv \frac{m}{4},
    \label{eq:momentum}\\
    P_{\varphi}=&-\frac{\partial R}{\partial \varphi}=n_1-n_2\equiv m_{12}, \label{eq:m12}\\
    P_{\theta}=&-\frac{\partial R}{\partial \theta}=n_3-n_4\equiv m_{34}. \label{eq:m34}
\end{align}

\begin{figure*}[t]
	\centering
	\scalebox{1.0}{\import{Pictures/}{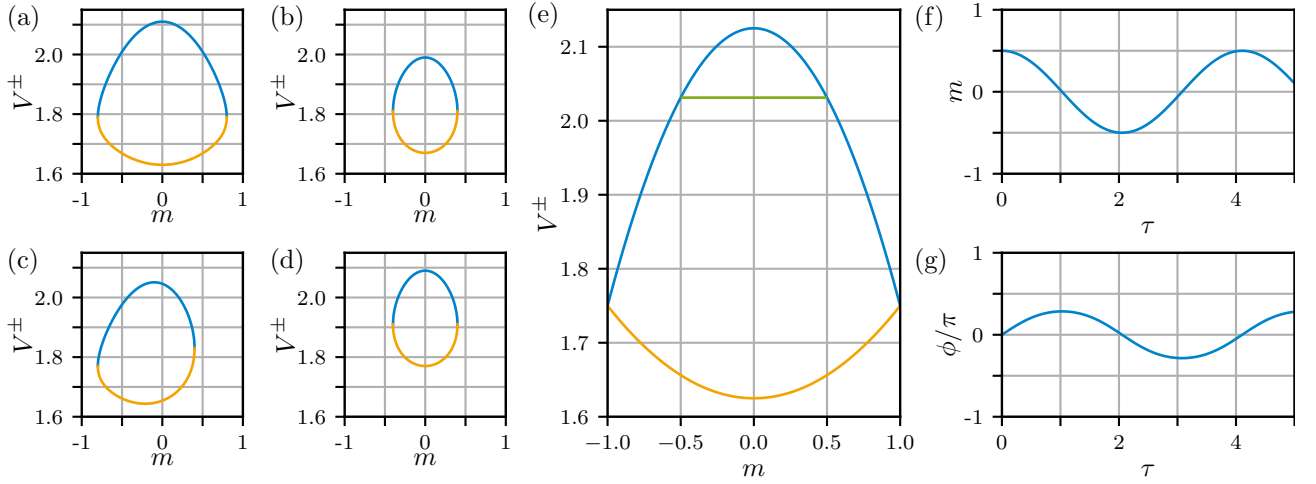}}
	\caption{(a)-(c) Potentials $V^+(m)$ (blue) and $V^-(m)$ (orange) versus population imbalance $m$ (\ref{eq:vPotentials}) for $\bar{\omega}_j=1$. For $m_{12}=m_{34}$ equal to ($0.1$ in (a), $0.3$ in (b)), the potentials are symmetric around the origin. Otherwise, this symmetry is broken ($m_{12}=0.1$, $m_{34}=0.3$ in (c)). (d) The potentials can be shifted in dimensionless energy by varying the value of the recoil frequencies ($\bar{\omega}_j=1.1$). (e) For $m_{12}=m_{34}=0$, $V^+$ and $V^-$ are symmetric in $m$ and $m\in[-1,1]$. For initial values $m_0=0.5$, $\phi_0=0$ and $\tau_0=0$, the dynamics of the Josephson variables described by (\ref{eq:analytical_m}) and (\ref{eq:analytical_phi}) are shown in (f) and (g). The energy is constant during the oscillation (green in (e)).}
	\label{fig:analytical_M}
\end{figure*}

In terms of the new coordinates the dimensionless Lagrangian $\mathcal{L}$ reads
\begin{align}
    \mathcal{L}=&\dot{\Phi}+\frac{m}{4}\dot{\phi}+m_{12}\dot{\varphi}+m_{34}\dot{\theta}-H(m,\phi),\label{eq:dimless_lagrangian}
\end{align}
with a generic Josephson Hamiltonian energy
\begin{gather}
    H(m,\phi)=\frac{\eta}{4}\cos\phi-\frac{m^2}{8}+
    \mathcal{C}, \label{eq:hamiltonian}\\
        \eta=\sqrt{
        	\left[(1+m)^2-4m^2_{12}\right]
        	\left[(1-m)^2-4m^2_{34}\right]}.
\end{gather}
Here, we have denoted an energy offset
$\mathcal{C}=
(m_{12}^2+m_{34}^2+2(\bar{\omega}_{12}m_{12}+\bar{\omega}_{34}m_{34})-
7/2+\sum_{j=1}^4\bar{\omega}_j)/4$ and transition energies
$\bar{\omega}_{12}=\bar{\omega}_1-\bar{\omega}_2$ and $\bar{\omega}_{34}=\bar{\omega}_3-\bar{\omega}_4$. As $\mathcal{L}$ does not depend on $\Phi$, $\varphi$ or $\theta$, these phases are cyclic \cite{Noether1918}. Therefore, the conjugate momenta, total particle number $N$ and population differences $m_{12}$ (\ref{eq:m12}), $m_{34}$ (\ref{eq:m34}) are conserved. Consequently, the equations of motion for $\Phi$, $\varphi$ and $\theta$ can be solved by quadrature.

Clearly $H$ (\ref{eq:hamiltonian}) is the Legendre transform of $\mathcal{L}$ (\ref{eq:dimless_lagrangian}). Accordingly, the dynamics of the system, using \eqref{eq:momentum} reads
\begin{align}
\begin{aligned}
    \dot{\phi}&=4\partial_{m}H=
    \cos{\phi}\partial_m\eta-m,\\ 
    \dot{m}&=-4\partial_{\phi}H=\eta\sin\phi.
\end{aligned}
    \label{eq:eomJosephson}
\end{align}
These are  Josephson-like differential equations \cite{leggett01,Smerzi1997a,Williams1999}.

\subsection{General solution}

In simple classical mechanics problems of particles with position $x$ and momentum $p$, Hamiltonian energies $H(x,p)=T(p)+V(x)$ separate into kinetic $T(p)$ and potential $V(x)$ energy. At the turning points $\dot{x}=\partial_p H=0$, the Hamilton function is purely determined by potential energy $H(x,p=0)=V(x)$. A similar investigation can be performed in the given case \cite{Braun1993,Raghavan1999}. Through a canonical transformation, we can exchange the role of position and momentum and consider $m$ as the position and $\phi$ as the momentum variable. Thus, at the turning points $\dot{m}=-4\partial_\phi H=0$, 
remarkably two momenta
    \begin{align}
        \phi^+=&0, & \phi^-=&\pi.
    \end{align}
    are possible. In turn, this defines two potentials
    \begin{align}
        H(m,\phi^\pm)=V^{\pm}(m)=\pm\frac{\eta}{4}
        -\frac{m^2}{8}
        +\mathcal{C}.\label{eq:vPotentials}
    \end{align}

Physical solutions with  energies $\varepsilon=H(m,\phi)$ must be constraint by these two potentials, $V^-<\varepsilon<V^+$. 
This limits the value range of $m$ and $\phi$ depending on the system parameters $m_{12}$ and $m_{34}$ (see Fig. \ref{fig:analytical_M}). As the energy of the system is conserved, the equation of motion (\ref{eq:eomJosephson}) for $m(\tau)$ can be expressed using the potentials $V^{\pm}$ as
    \begin{align}
        \dot{m}=\pm 4\sqrt{(V^+(m)-\varepsilon)(\varepsilon-V^-(m))}.
    \end{align}
Thus, the dynamical solution $\tau(m)$ can be calculated as
    \begin{align}
        \tau(m)-\tau_0=\int_{m_0}^{m}\frac{\pm \dd \zeta}{4\sqrt{(V^+(\zeta)-\varepsilon)(\varepsilon-V^-(\zeta))}}.
    \end{align}
This relation can be inverted piecewise to obtain $m(\tau)$.

\subsection{Analytical solution for $m_{12}=m_{34}=0$}

For the special case $m_{12}=m_{34}=0$, implying $n_1=n_2$ and $n_3=n_4$, an analytical expression for the dynamical solution $m(\tau)$ can be given in terms of the elliptic cosine $\text{cn}(u)$ \cite{DLMF} as
    \begin{align}
        m(\tau)=\pm&\sqrt{\frac{\mu+2}{3}}\;\text{cn}\left(\xi(\tau-\tau_0),\rho^2\right), \label{eq:analytical_m}
    \end{align}
where $\mu=m_0^2+2(m_0^2-1)\cos\phi_0$, $\xi=\sqrt{6-3\mu}/2$ and $\rho^2=(\mu+2)/(6-3\mu)$. With that, the dynamical solution of the phase $\phi(\tau)$ can be calculated by integration of (\ref{eq:eomJosephson}), yielding
\begin{align}
    \phi(\tau)=&2\,\text{arctan}\bigl\{\sqrt{3}\,\text{tanh}\bigl[ \ln(1-\rho) \nonumber\\
    &-\ln(
    \text{dn}\left(\xi(\tau-\tau_0),\rho^2\right)-\rho\,\text{cn}\left(\xi(\tau-\tau_0),\rho^2\right))\nonumber\\
    &+\text{arctanh}\bigl(\text{tan}(\phi_0/2)/\sqrt{3}\bigr)\bigr]\bigr\},
    \label{eq:analytical_phi}
\end{align}
with the delta ampltiude $\text{dn}(u)$ \cite{DLMF}. The analytical solutions for $m(\tau)$ and $\phi(\tau)$ as well as visualizations of the potentials $V^+$ and $V^-$ can be seen in Fig. \ref{fig:analytical_M}.

\begin{figure}[t]
 	\centering
 	\scalebox{1.0}{\import{Pictures/}{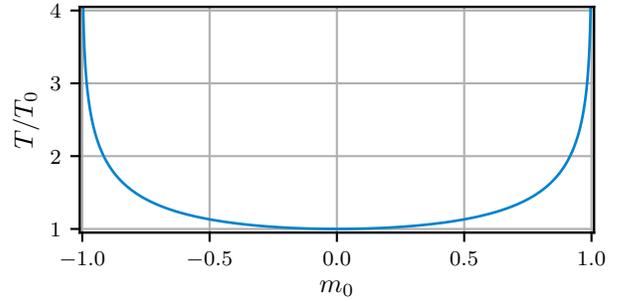}}
 	\caption{Period of FWM oscillation $T$ versus initial population imbalance $m_0$, normed to $T_0$. The period diverges when nearing aperiodic solutions.}
 	\label{fig:period_4WM}
\end{figure}

The period of the motion can be calculated as
    \begin{align}
    \label{eq:period}
        T=\frac{4\;\text{K}(\rho^2)}{\xi}.
    \end{align}
There, K is the complete elliptic integral of first kind \cite{DLMF}. The basic frequency of the oscillation $T_0=T(m_0=0)$ can be calculated as
    \begin{align}
        T_0=4\pi/\sqrt{12}.
    \end{align}
As can be seen in Fig. \ref{fig:period_4WM}, the period of the FWM oscillation diverges when nearing the regime of aperiodic solutions.

\section{Four-wave mixing with background population} \label{sec:background}

In the ideal FWM setting, the residual wave $\ket{\psi_\beta}$ is absent. However, additional momentum states might be populated accidentally during the initialization procedure or system evolution. To investigate this scenario, we simulate the dynamics of the system, described by the Gross-Pitaevskii equation (\ref{eq:GPE}), using a Runge-Kutta scheme and Fast Fourier Transforms (FFT) on a discrete periodic lattice. We use a two-dimensional lattice with $16\times 16$ sites while setting $\gamma=1/$s and discretizing dimensionless time with $\Delta\tau=10^{-6}$. For implementation we choose the geometry of FWM states described in App. \ref{sec:state_preparation}, yielding $\bar{\omega}_j=1$ for $j=1,\dots,4$. The populations are set to $n_1=n_2=0.375$ and $n_3=n_4=0.125$, resulting in $m_{12}=m_{34}=0$ and $m_0=0.5$. All phases are set to $\varphi_j=0$, yielding $\phi_0=0$.

\begin{figure}[t]
 	\centering
 	\scalebox{1.0}{\import{Pictures/}{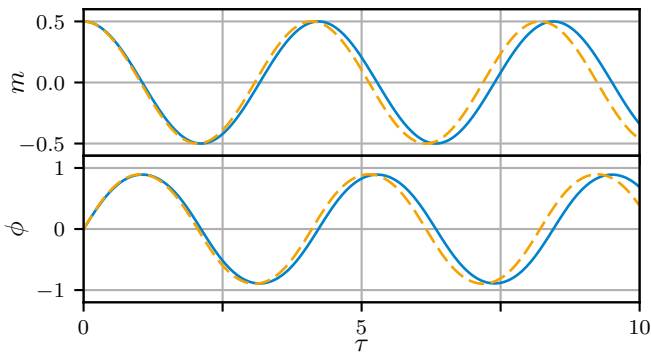}}
 	\caption{Population imbalance $m(\tau) $ and relative phase $\phi(\tau)$ versus dimensionless time $\tau$ for analytical (orange, dashed) and numerical GP simulation on a discrete periodic lattice (blue, solid).}
 	\label{fig:background_ana_mws}
\end{figure}

\begin{figure}[b]
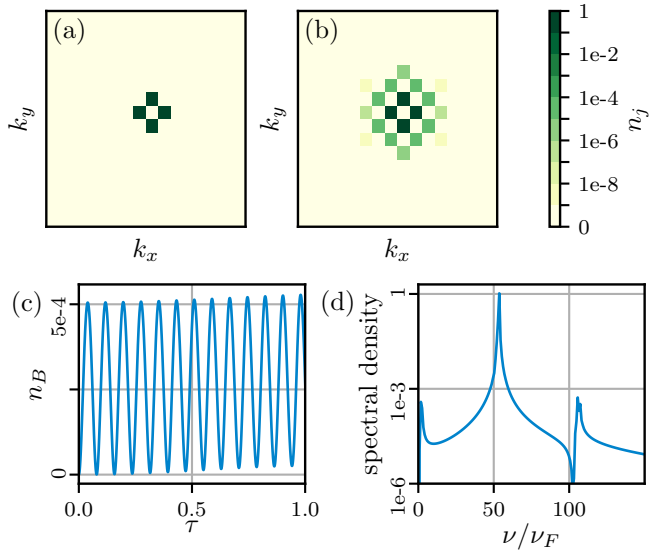

    \begin{subfigure}
        \centering
        \scalebox{1.0}{\import{Pictures/}{background_histogram_1.pgf}}
    \end{subfigure}
    \begin{subfigure}
         \centering
        \scalebox{1.0}{\import{Pictures/}{background_histogram_2.pgf}}
    \end{subfigure}
 	\caption{Histograms of populations on discrete $16\times 16$ lattice in $k_x$-$k_y$-plane at $\tau=0$ (a) and $\tau=5.0$ (b). (c) Background population $n_B$ starts oscillating and quickly reaches maximum value. (d) Oscillation frequency of $n_B$ is about 50 times bigger than the FWM frequency $\nu_F$.}
 	\label{fig:background_histogram}
\end{figure}

As can be seen in Fig. \ref{fig:background_ana_mws}, the numerical results of the GP simulation start to deviate from the four-mode approximation \eqref{eq:complexdgl} already after a few cycles, noticeably. Looking at $m(\tau)$ and $\phi(\tau)$, the numerical results show a larger period of the oscillation. However, the general shape of the oscillations remains unchanged.

This behaviour is caused by an instability of the simulation due to numerical noise of the FFT producing population on the grid outside of the FWM states. As depicted in Fig. \ref{fig:background_histogram} (a), the system is prepared at $\tau=0$ with population only present in the FWM states. However, the histogram in Fig. \ref{fig:background_histogram} (b) at $\tau=5$ clearly shows that additional states in the vicinity of the FWM states have been populated. As this background population is located at the center of the lattice, the chosen grid is large enough such that no edge effects occur during the simulation.

\begin{figure}[b]
 	\centering
 	\scalebox{1.0}{\import{Pictures/}{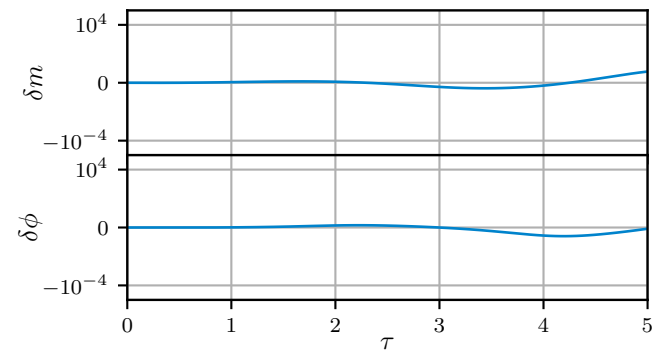}}
 	\caption{Deviations $\delta m=(m_n-m_a)/\text{max}(m_a)$ and $\delta \phi=(\phi_n-\phi_a)/\text{max}(\phi_a)$ versus dimensionless time $\tau$ from numerical ($m_n$, $\phi_n$) to analytical solutions ($m_a$, $\phi_a$). All population outside of FWM states is eliminated after each simulation step.}
 	\label{fig:ana_mws_mask}
\end{figure}

Yet, the instability caused by accidental population of additional momentum states is not destructive in nature. Looking at Fig. \ref{fig:background_histogram} (c), the total background population
    \begin{align}
        n_B=\sum_{l>4}\vert\beta_l\vert^2
    \end{align}
grows rapidly at the beginning of the oscillation. Subsequently, the dynamics of $n_B(\tau)$ stabilize and show oscillations with a maximum value of around $n_B\simeq 5\cdot 10^{-4}$. As can be seen in Fig. \ref{fig:background_histogram} (d), the frequency of the ensuing oscillation is about $50$ times larger than the FWM frequency
    \begin{align}
        \nu_{F}=\frac{1}{T(m_0=0.5)}\simeq 0.244.
    \end{align}

The non-negligible background population is the cause of change in the dynamics of the FWM process. Because of
    \begin{align}
        n_{F}+n_B=&1, & n_{F}=&\sum_{j=1}^4\vert\alpha_j\vert^2,
    \end{align}
growing $n_B$ reduces the population in the FWM states $n_F$ in comparison to the ideal case. As the FWM process is caused by the density-density-interaction terms in the Gross-Pitaevskii equation (\ref{eq:GPE}), even small changes in the particle number participating in the process have profound effects on the dynamics.

The analytical solution can be recovered by eliminating all numerical noise produced by FFT after each simulation step. Using such masks in $k$-space, the numerical simulation and analytical solution agree within about $10^{-5}$ (see Fig. \ref{fig:ana_mws_mask}). However, this procedure yields a loss in total particle number of about $\Delta N/N=10^{-6}$, far surpassing typical numerical noise.

For the implementation of the FWM neuron, we are interested in rather short time scales and more qualitative behaviour of the system. Therefore, we accept the change in frequency of the FWM oscillations and use the simulation on a discrete periodic lattice in the investigations without additionally applying a filter mask in $k$-space. This is beneficial due to the high flexibility of the simulation regarding initial conditions of the FWM states. However, the deviation between the ideal case and with present background population should be kept in mind, especially when looking at increasing simulation times.

\section{Four-wave mixing neuron} \label{sec:4WM_neuron}

Artificial neurons are the basic computation units in neural networks \cite{Rosenblatt1958}. In addition to the processing of real numbers, such systems are also able to operate with complex-valued inputs and outputs \cite{Hirose2012}. As the FWM process is described in terms of complex amplitudes $\alpha_j$, the presented implementation of the FWM neuron constitutes a complex-valued neuron. Due to the experimental accessibility of particle numbers and phases, we choose to describe the nonlinear activation function and the learning process in terms of absolute values and phases, rather than using real and imaginary parts of the complex amplitudes $\alpha_j$ \cite{Hirose2012}.

\begin{figure}[t]
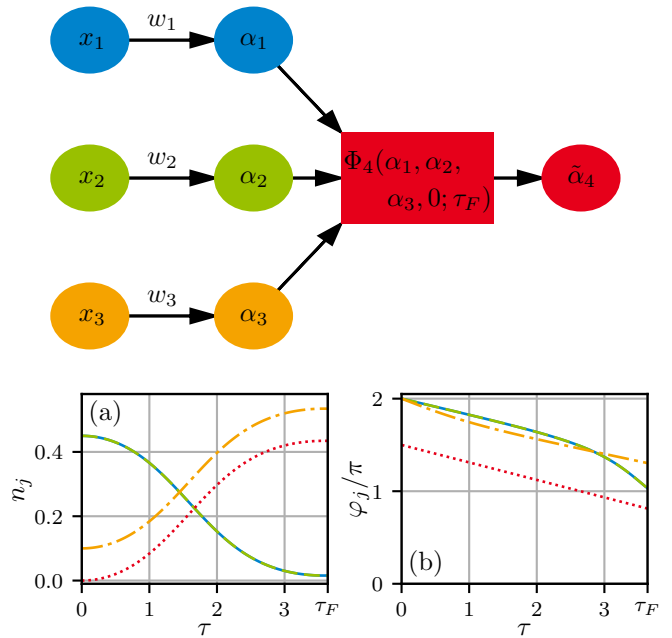

	\begin{subfigure}
		\centering
		\scalebox{1.0}{\import{Pictures/}{neuron_algorithm_1.pgf}}
	\end{subfigure}
	\begin{subfigure}
		\centering
		\scalebox{1.0}{\import{Pictures/}{neuron_algorithm_2.pgf}}
	\end{subfigure}
	\caption{Initialisation sequence for a FWM neuron.
	Classical inputs $x_j$ are weighted with $w_j$, yielding amplitudes $\alpha_j$. The nonlinear relation $\Phi_4(\alpha_1,\alpha_2,\alpha_3,0;\tau_F)$ yields output $\tilde{\alpha}_4$. 
	The duration $\tau_F=T/2$ is determined for $n_1=n_2=0.45$ (blue, solid; green, dashed) and $n_3=0.1$ (orange, dash-dotted; $n_4$: red, dotted), while $\varphi_j=0$, as a half-oscillation period leading to maximal response.}
	\label{fig:neuron_algorithm}
\end{figure}

In general, complex-valued artificial neurons process an $n$-dimensional input $x_j=\vert x_j\vert e^{i\kappa_j}$, $j=1,\dots, n$, by multiplying individually with weights $w_j=\vert w_j\vert e^{i\vartheta_j}$, summing up the weighted inputs $v_j=w_jx_j$ and yielding an output $y$ via a nonlinear activation function $\Omega$,
\begin{align}
    y=&\Omega(u), & u=\sum_{j=1}^n v_j. \label{eq:naf_neuron}
\end{align}

We implement such a computational unit with the FWM process on coherent matter-waves. The phase-flow 
   \begin{align}
        \tilde{\boldsymbol{\alpha}}=
        \Phi(\boldsymbol{\alpha};\tau_F),
    \end{align}
maps the initial state $\boldsymbol{\alpha}=
(\alpha_1,\dots,\alpha_4)$
to the evolved state   $\tilde{\boldsymbol{\alpha}}=(\tilde{\alpha}_1,\dots,\tilde{\alpha}_4)$ after 
the duration  $\tau_F$ of the FWM process \eqref{eq:complexdgl}. Identifying the three amplitudes $\alpha_1$, $\alpha_2$ and $\alpha_3$ as weighted inputs $v_j$ and $\tilde{\alpha}_4$ as output $y$, a similar, though not identical, rule to (\ref{eq:naf_neuron}) can be established
    \begin{align}
        \tilde{\alpha}_4=\Phi_4(\alpha_1,\alpha_2,\alpha_3,0;\tau_F).
    \end{align}
The fourth component of the phase-flow map constitutes a nonlinear activation function of a complex-valued FWM neuron with three input channels. In an experiment, we use externally stored weights $w_j$ for the neuron, the classical input data $x_j$ and prepare the weighted input amplitude
\begin{align}
\label{eq:complexweigthedinput}
    \alpha_j=w_jx_j
\end{align}
by  a sequence of Bragg pulses (see App. \ref{sec:state_preparation}).

\begin{figure*}[t]
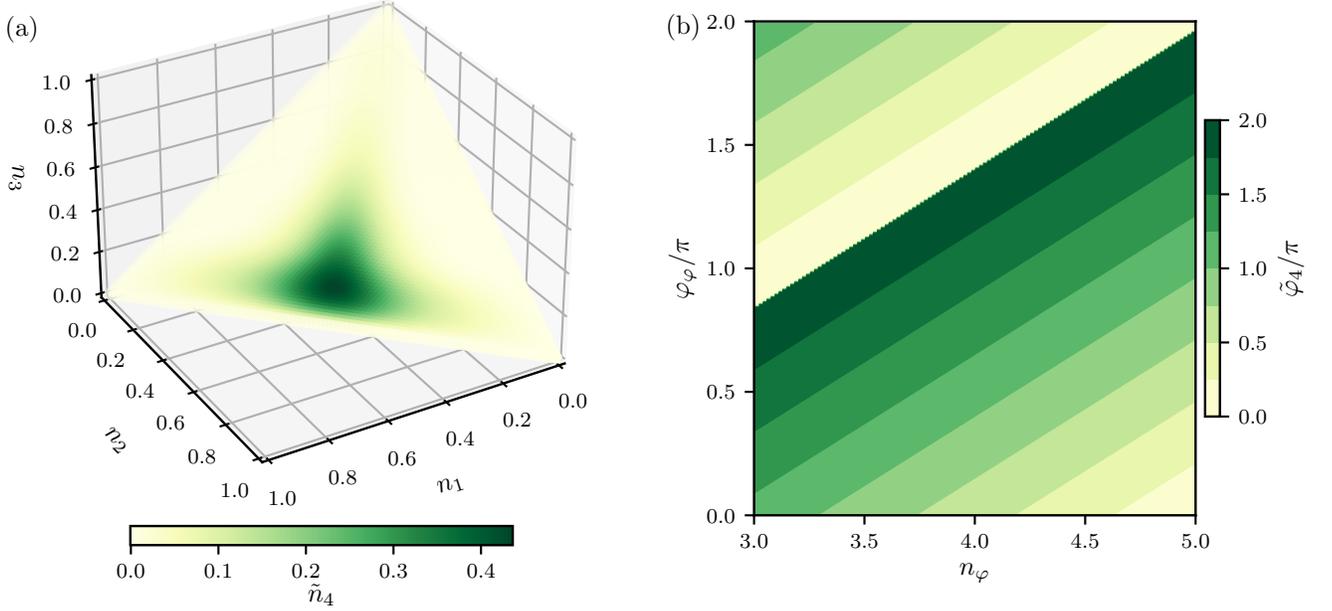

    \begin{subfigure}
        \centering
        \scalebox{1.0}{\import{Pictures/}{nonlinear_activation_function_N.pgf}}
    \end{subfigure}
    \begin{subfigure}
         \centering
        \scalebox{1.0}{\import{Pictures/}{nonlinear_activation_function_phi.pgf}}
    \end{subfigure}
 	\caption{Nonlinear activation function $\Phi_4(\alpha_1,\alpha_2,\alpha_3,0;\tau_F)$ of the FWM neuron in terms of (a) $\tilde{n}_4$ (\ref{eq:naf_fwm_n}) versus $n_1$, $n_2$ and $n_3$ and (b) $\tilde{\varphi}_4$ (\ref{eq:naf_fwm_phi}) versus $n_{\varphi}$ and $\varphi_{\varphi}$ (\ref{eq:phase_linear_combinations}).}
 	\label{fig:nonlinear_activation_function}
\end{figure*}

\subsection{Nonlinear activation function}

In order to quantify the nonlinear activation function, $\tau_F$ has to be determined. To do so, we choose $n_1=n_2=0.45$ and $n_3=0.1$, while setting $\varphi_j=0$. The resulting FWM oscillation can be seen in Fig. \ref{fig:neuron_algorithm}. To maximize the output in terms of $\tilde{n}_4$ for this scenario, we set
    \begin{align}
        \tau_F=T/2,
    \end{align}
where $T$ is the oscillation period as in 
\eqref{eq:period}.

The FWM neurons response is 
calculated for varying weighted inputs numerically
(cf. Sec.~\ref{sec:background}).
We tune $n_j$ from $0$ to $1$ subject to the constraint $\sum_{j=1}^4n_j=1$. 
Due to probability (number) conservation, all admissible combinations of $n_j$ form a plane in $n_1$-$n_2$-$n_3$-space. 
The input phases $\phi_j$ are varied from $0$ to $2\pi$.  
The results can be seen in Fig. \ref{fig:nonlinear_activation_function}. The output particle number 
    \begin{align}
        \tilde{n}_4=\vert\Phi_4(\alpha_1,\alpha_2,\alpha_3,0;\tau_F)\vert^2 \label{eq:naf_fwm_n}
    \end{align}
is independent of the input phases $\varphi_j$. Hence, only the input particle numbers $n_j$ determine this part of the output. While there is no analytical expression for the relation, it can be extracted from Fig. \ref{fig:nonlinear_activation_function}, that there has to be an exchange symmetry regarding $n_1$ and $n_2$.

The output phase
    \begin{align}
        \tilde{\varphi}_4=\text{arg}\left[\Phi_4(\alpha_1,\alpha_2,\alpha_3,0;\tau_F)\right] \label{eq:naf_fwm_phi}
    \end{align}
exhibits a remarkable simple behaviour. 
By analyzing Fig.~\ref{fig:nonlinear_activation_function} (b), we find 
    \begin{align}
        n_{\varphi}=&3n_1+3n_2+5n_3, & \varphi_{\varphi}=&\varphi_1+\varphi_2-\varphi_3. \label{eq:phase_linear_combinations}
    \end{align}
Accordingly, the input-output-relation reads
\begin{align}
    \tilde{\varphi}_4=
    s n_{\varphi}+\varphi_{\varphi}+d, \label{eq:phase_output}
\end{align}
where the slope and offset of phase were
determined from a fit as $s=(-1.77\pm0.01)$ and $d=(2.67\pm0.04)$.

The numerical results in Fig. \ref{fig:nonlinear_activation_function} can be used to determine the partial derivatives $\partial \tilde{n}_4/\partial n_j$, $\partial \tilde{\varphi}_4/\partial n_j$ and $\partial \tilde{\varphi}_4/\partial \varphi_j$. These are needed to be able to train the neuron according to a steepest descent method.

\begin{figure*}[t]
	\centering
	\scalebox{1.0}{\import{Pictures/}{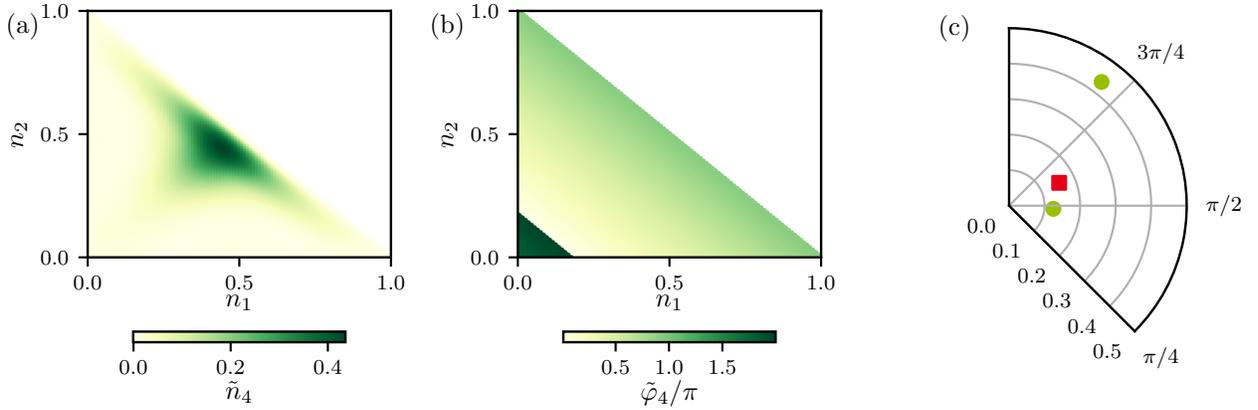}}
	\caption{Input-output relations (a) $\tilde{n}_4(n_1,n_2)$ and (b) $\tilde{\varphi}_4(n_1,n_2,0,0)$ of the FWM neuron to solve the XOR problem. (c) By choosing the outputs of the individual cases according to Tab.~\ref{tab:xor_4WM} (green, output $0$; red, output $1$), the XOR problem is solvable using a single FWM neuron.}
	\label{fig:xor_4WM_neuron}
\end{figure*}

\subsection{Steepest descent learning for complex-valued neurons}

Steepest descent methods are common procedures in optimization, as well as in supervised learning in neural networks \cite{Goodfellow2016}. We consider the case of a single output neuron. In so-called error-correction learning, this neuron is stimulated by an input vector $\vec{x}^{(i)}=(x_1^{(i)},x_2^{(i)},x_3^{(i)})$, where $i$ denotes an instant in time at which the excitation is applied to the system. The training dataset is described by
    \begin{align}
        \mathcal{T}:\left\{ \vec{x}^{(i)},\hat{\alpha}_4^{(i)};i=1,\dots,\mathcal{M}\right\},
    \end{align}
where $\hat{\alpha}_4^{(i)}$ is the desired response associated with $\vec{x}^{(i)}$ and $\mathcal{M}$ is the size of the dataset. In response to this stimulus, the neuron produces an output $\tilde{\alpha}_4^{(i)}$.

Starting from initial weights $\vec{w}=(w_{1},\dots,w_n)$, the goal of the learning procedure is to adjust the weights to minimize the difference between the desired and actual outputs, described by means of a cost function $\mathcal{F}$. A typical cost function is the squared error averaged over the training sample set \cite{Bassey2021}
    \begin{align}
        \mathcal{F}&=
        \frac{1}{\mathcal{M}}\sum_{i=1}^{\mathcal{M}} \mathcal{F}^{(i)}, &
         \mathcal{F}^{(i)}&=\frac{1}{2}\left\vert \tilde{\alpha}_4^{(i)}-\hat{\alpha}_4^{(i)}\right\vert^2.
        \label{eq:averaged_error_energy}
    \end{align}

This yields an unconstrained optimization problem with the necessary condition for optimality $\nabla\mathcal{F}=0$, where $\nabla$ denotes the gradient operator in weight space. In a steepest descent method, adjustments applied to the weight vector are performed in the direction of the negative gradient
    \begin{align}
        \Delta \vec{w}(n)=\vec{w}(n+1)-\vec{w}(n)=-\lambda\nabla\mathcal{F},
    \end{align}
where $n$ symbolizes one iteration and $\lambda$ is a positive learning rate.

In the on-line learning approximation \cite{Haykin2009}, adjustments to the weights are performed on an example-by-example basis. The cost function to minimize is therefore the instantaneous error energy $\mathcal{F}^{(i)}$. An epoch consists of $\mathcal{M}$ training samples. At an instant $i$, a pair $\{\vec{x}^{(i)},\hat{\alpha}_4^{(i)}\}$ is presented to the neuron and weight adjustments are performed. Subsequently, the next sample is presented to the network until all $\mathcal{M}$ samples have been evaluated.

The absolute values and phases of the weights can be updated independently \cite{Hirose2012}
\begin{align}
    \Delta\vert w_j^{(i)}\vert=&-\lambda_a\partial_{\vert w_j\vert}\mathcal{F}^{(i)}, & \Delta\vartheta_j^{(i)}=-\lambda_p\partial_{\vartheta_j}\mathcal{F}^{(i)}, \label{eq:update_rules}
\end{align}
where $\lambda_a$ and $\lambda_p$ are the learning rates for absolute value and phase respectively. The required gradients for the update rules (\ref{eq:update_rules}), keeping in mind the variable dependencies of the nonlinear activation function, are calculated using the chain rule as
    \begin{align}
        \begin{aligned}
        \frac{\partial \tilde{n}_4}{\partial\vert w_j\vert}=&\frac{\partial \tilde{n}_4}{\partial n_j}\frac{\partial n_j}{\partial\vert w_j\vert}=\vert x_j\vert\frac{\partial \tilde{n}_4}{\partial n_j}, \\
        \frac{\partial \tilde{\varphi}_4}{\partial\vert w_j\vert}=&\frac{\partial \tilde{\varphi}_4}{\partial n_j}\frac{\partial n_j}{\partial\vert w_j\vert}= \vert x_j\vert\frac{\partial \tilde{\varphi}_4}{\partial n_j}, \\
        \frac{\partial \tilde{\varphi}_4}{\partial \vartheta_j}=&\frac{\partial \tilde{\varphi}_4}{\partial \varphi_j}\frac{\partial \varphi_j}{\partial \vartheta_j}=\frac{\partial \tilde{\varphi}_4}{\partial \varphi_j}.
        \end{aligned}
    \end{align}
Hence, the update rules for $\vert w_j\vert$ and $\vartheta_j$ are
    \begin{align}
        \Delta\vert w_j^{(i)}\vert=&-\lambda_a \Bigl[\left(\tilde{n}_4^{(i)}-\hat{n}_4^{(i)}\cos(\tilde{\varphi}_4^{(i)}-\hat{\varphi}_4^{(i)})\right)\partial_{n_j}\tilde{n}_4^{(i)} \nonumber \\
        &+\tilde{n}_4^{(i)}\sin(\tilde{\varphi}_4^{(i)}-\hat{\varphi}_4^{(i)})\partial_{n_j}\tilde{\varphi}_4^{(i)}\Bigr]\vert x_j^{(i)}\vert, \label{eq:update_weight_abs}\\
    \Delta\vartheta_j^{(i)}=&-\lambda_p \tilde{n}_4^{(i)}\hat{n}_4^{(i)}\sin(\tilde{\varphi}_4^{(i)}-\hat{\varphi}_4^{(i)})\partial_{\varphi_j}\tilde{\varphi}_4^{(i)}.
    \label{eq:update_weight_phase}   
\end{align}

\subsection{Application: XOR problem}

\begin{table}[t]
\begin{center}
\begin{tabular}{ |c c | c| }
\hline
Input 1 & Input 2 & Output \\
\hline
0 & 0 & 0 \\
0 & 1 & 1 \\
1 & 0 & 1 \\
1 & 1 & 0 \\
\hline
\end{tabular}
\end{center}
\caption{Input-output mapping for the XOR problem.}
\label{tab:xor}
\end{table}

To investigate the calculation and learning abilities of the FWM neuron, we use it to solve the XOR problem. The input-output mapping for this problem is shown in Tab.~\ref{tab:xor}. The XOR problem consists of two real-valued binary inputs. The output is supposed to be $0$ if the two inputs are identical and $1$ if they are different. It has been shown, that this problem is not solvable for a single real-valued neuron, i.e. hidden layers are required \cite{Minsky1969}. However, a single complex-valued neuron is able to solve this problem \cite{Nitta2003}.

\subsubsection{Input and output encoding}

\begin{table}[b]
\begin{center}
\begin{tabular}{ |c c|c c ||c | c c| }
\hline
Input 1 & Input 2 &$\vert x_1\vert$ & $\vert x_2\vert$ & Output & $\tilde{n}_4$ & $\tilde{\varphi}_4$ \\
\hline
0 & 0 & 0.3 & 0.3 & 0 & 0.125 & 1.5 \\
0 & 1 & 0.3 & 0.45 & 1 & 0.155 & 2.0 \\
1 & 0 & 0.45 & 0.3 & 1 & 0.155 & 2.0 \\
1 & 1 & 0.45 & 0.45 & 0 & 0.435 & 2.5\\
\hline
\end{tabular}
\end{center}
\caption{Encoded input-output mapping for the XOR problem using the FWM neuron.}
\label{tab:xor_4WM}
\end{table}

To use the full value range of the nonlinear activation function of the FWM neuron to solve the XOR problem, an encoding scheme for the inputs and the output has to be developed. The inputs $x_{1,2}$ are chosen to lie on the positive real axis ($\kappa_j=0$). While an input $0$ is identified by $\vert x_j\vert=0.3$, an input $1$ is given by $\vert x_j\vert=0.45$.

The weights $w_j$ of the neuron are still allowed to possess non-vanishing phases $\vartheta_j$. Therefore, the weighted inputs presented to the FWM neuron will be given by
    \begin{align}
        \sqrt{n_j}=&\vert w_j\vert\vert x_j\vert, & \varphi_j=&\vartheta_j.
    \end{align}
As two input particle numbers, chosen to be $n_1$ and $n_2$, of the FWM neuron are set using this encoding, the third, in this case $n_3$, is automatically determined to ensure $\sum_jn_j=1$. Consequently, the combinations of inputs $n_1$ and $n_2$ are constrained by $0\leq n_1+n_2\leq 1$.

The particle number response of the FWM neuron to the inputs is completely determined by the input particle numbers $\tilde{n}_4(n_1,n_2)$. The neuron response in terms of the phase follows
    \begin{align}
        \tilde{\varphi}_4(n_1,n_2,\varphi_1,\varphi_2)=\tilde{\varphi}_4(n_1,n_2,0,0)+\varphi_1+\varphi_2
    \end{align}
These input-output relations can be seen in Fig. \ref{fig:xor_4WM_neuron}.

The possible outputs of the XOR problem are encoded in a similar fashion. An output $0$ is encoded via $\tilde{n}_4=0.125$ and $\tilde{\varphi}_4=1.5$ or $\tilde{n}_4=0.435$ and $\tilde{\varphi}_4=2.5$ for the input cases $[0,0]$ and $[1,1]$ respectively. The output $1$ is always encoded as $\tilde{n}_4=0.155$ and $\tilde{\varphi}_4=2$. The complete encoding of the XOR problem for the FWM neuron can be seen in Tab.~\ref{tab:xor_4WM}. The presented encoding is completely equivalent to the original XOR problem. Hence, it can be used to solve the problem by means of the FWM neuron.

\subsubsection{Training results}

Starting from random initial weights, the update rules (\ref{eq:update_weight_abs}) and (\ref{eq:update_weight_phase}) are used to train the FWM neuron to solve the XOR problem. Training epochs are performed with $\mathcal{M}=1000$ random samples. The learning rate of the phase $\lambda_p=10^{-8}$ is kept constant for all epochs while the absolute value learning rate $\lambda_a$ is gradually reduced from $10^{-3}$ to $10^{-4}$ during the training. After each epoch, the performance of the neuron is evaluated by calculating the averaged squared error $\mathcal{F}$ according to (\ref{eq:averaged_error_energy}) using all four possible input-output pairs of the XOR problem.

\begin{figure}[b]
 	\centering
 	\scalebox{1.0}{\import{Pictures/}{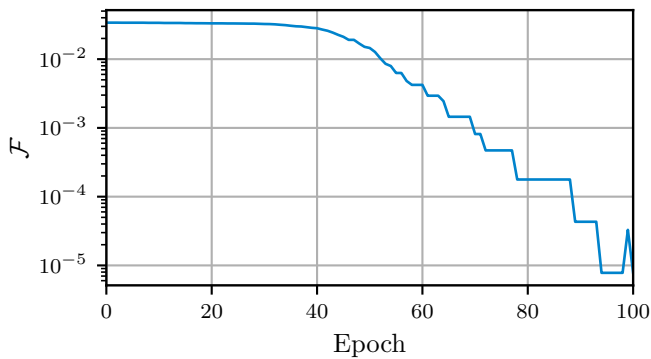}}
 	\caption{Averaged squared error $\mathcal{F}$ (\ref{eq:averaged_error_energy}) over all four possible input-output pairs of the XOR problem versus number of training epochs.}
 	\label{fig:training_xor}
\end{figure}

As can be seen in Fig. \ref{fig:training_xor}, the FWM neuron is able to learn to solve the XOR problem. After $100$ training epochs, the initial error is reduced to $\mathcal{F}=7.8\cdot 10^{-6}$. A sample is categorized as being identified correctly, if the neuron output is within $\pm 0.005$ in terms of particle number and within $\pm 0.05$ in terms of phase of the desired value. At the end of the training procedure, every test sample is identified correctly.

\section{Conclusion \& Perspectives}

We investigated the ideal FWM process in a three-dimensional homogeneous BEC. By introducing appropriate coordinates, we showed, that the dynamics of the system exhibit Josephson-like oscillations, which can be described analytically by means of elliptic functions. These analytical expressions agree with numerical simulations of the Gross-Pitaevskii equation on a discrete periodic lattice. We investigated the influence of additional population outside of the FWM states on these dynamics. While the frequency of the oscillations changes, the main characteristics of the dynamics persist.

Identifying three complex amplitudes of the FWM setup as input and the fourth amplitude as output, we introduced a new implementation for a complex-valued artificial neuron. We investigated the nonlinear activation function of the FWM neuron and showed its learning capabilities using steepest descent learning for complex-valued neurons. These are demonstrated by solving the XOR problem using the FWM neuron. After completing $100$ learning epochs, the FWM neuron was able to identify every test sample presented to it correctly.

Looking ahead, we aim to implement the FWM neuron in a deep neural network. For this, two key aspects have to be investigated: parallelization ability and communication between layers of the network. Preliminary investigations showed, that multiple FWM neurons can be run in parallel by stacking the FWM setup in momentum space. Furthermore, light sheets can be used to stack multiple FWM neurons in real space and run them in parallel. However, more effort has to be put into investigations whether and how such stacked FWM neurons influence each other. Transporting information through a FWM neural network would require a delicate sequence of Bragg pulses to initialize output of one layer as input for the subsequent one. The exact nature of these pulse sequence has to be developed and investigated in detail.

\section*{Acknowledgments}

This work is supported by the DLR German Aerospace Center with funds provided by the Federal Ministry for Economic Affairs and Energy (BMWi) under Grant No. 50WM2250E.

\appendix
\section{Four-wave mixing state preparation}\label{sec:state_preparation}

The desired state after initialization for FWM is a superposition of plane waves with wave vectors $\vec{k}_1$, $\vec{k}_2$, $\vec{k}_3$ and $\vec{k}_4$, fulfilling the conditions (\ref{eq:4WM_cond}). There, all possible combinations for populations $n_1$, $n_2$, $n_3$ and $n_4$ with $\sum_{j=1}^4n_j=1$ should be realizable.

\begin{figure}[t]
	\centering
	\scalebox{1.0}{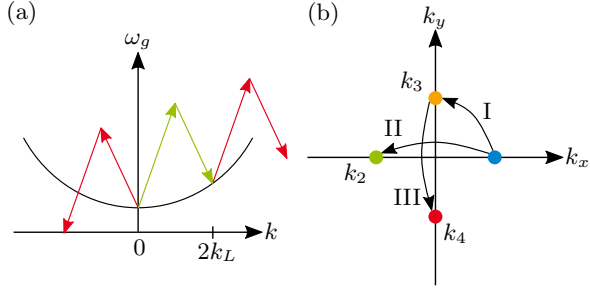}
	\caption{(a) Energy diagram for Bragg-diffraction versus wave number $k$. A ground-state BEC initially at rest experiences population transfer to a state with $2\vec{k}_L$ (green). Population transfer to other momentum states do not appear (red), as the conditions (\ref{eq:conditions_bragg}) are not fulfilled. (b) Proposed initialization sequence for FWM setup fulfilling (\ref{eq:4WM_cond}). Three Bragg pulses are used to set up any combination of populations between the FWM states, while ensuring that no population is transferred outside of the FWM states.}
	\label{fig:bragg_diffraction}
\end{figure}

\begin{figure}[t]
	\centering
	\scalebox{1.0}{\import{Pictures/}{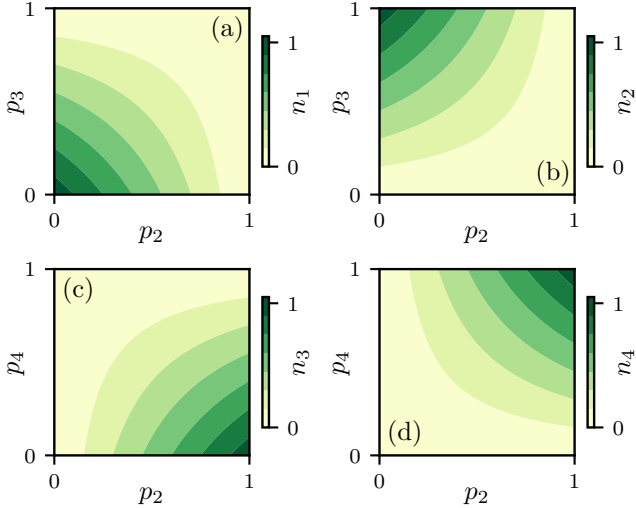}}
	\caption{Resulting particle numbers (a) $n_1$, (b) $n_2$, (c) $n_3$ and (d) $n_4$ for the initialization sequence described by Fig. \ref{fig:bragg_diffraction} and Table \ref{tab:preparation_sequence}. 
	All probabilities range from $0$ to $1$, conserving the total probability (particle number) \eqref{eq:number_conservation_preparation}.}
	\label{fig:preparation_sequence}
\end{figure}

\begin{table}[b]
\begin{center}
\begin{tabular}{ |c c c c c c| }
\hline
pulse & $n_0$ & $n_1$ & $n_2$ & $n_3$ & $n_4$ \\
\hline
0 & 0 & 1 & 0 & 0 & 0 \\
I & 0 & $(1-p_2)$ & 0 & $p_2$ & 0 \\
II & 0 & $(1-p_3)(1-p_2)$ & $p_3(1-p_2)$ & $p_2$ & 0 \\
III & 0 & $(1-p_3)(1-p_2)$ & $p_3(1-p_2)$ & $(1-p_4)p_2$ & $p_4p_2$\\
\hline
\end{tabular}
\end{center}
\caption{Particle numbers in the FWM states after each Bragg pulse of the sequence described in Fig.~\ref{fig:bragg_diffraction}.}
\label{tab:preparation_sequence}
\end{table}

We suggest to use atomic beamsplitters based on Bragg diffraction to populate the momentum states. This method is based on the interaction between the BEC in its internal ground state and two counterpropagating laser beams. In this scenario, energy and momentum conservation have to hold \cite{Neumann2021},
\begin{align}
    \hbar\omega_1+\frac{\hbar^2k_i^2}{2m}=&\hbar\omega_2+\frac{\hbar^2k_f^2}{2m}, & \vec{k}_i+\vec{k}_1=\vec{k}_f+\vec{k}_2, \label{eq:conditions_bragg}
\end{align}
with the initial $\vec{k}_i$ and the final $\vec{k}_f$ wave vector of the BEC and the frequencies $\omega_1$ and $\omega_2$ and wave vectors $\vec{k}_1$ and $\vec{k}_2$ of the laser beams respectively. If the two laser beams are perfectly anti-collinear, the momentum transfer in the BEC can be characterized as
\begin{align}
    \vec{k}_f-\vec{k}_i=\vec{k}_1-\vec{k}_2=2\vec{k}_L,
\end{align}
where $\vec{k}_L=(\vec{k}_1+\vec{k}_2)/2$.

For a shallow lattice ($U(\vec{r})=0$), the ground state energy $\hbar\omega_g$ of the BEC scales quadratically with the wave number, $\omega_g\propto k^2$. Hence, the laser frequencies have to be chosen carefully, such that population transfer between momentum states is energetically permitted (see Fig. \ref{fig:bragg_diffraction}).

A controlled initialization of momentum states can be performed, as initial states can be targeted individually and final states are given by the momentum and energy conditions (\ref{eq:conditions_bragg}). In Bragg diffraction, the portion of the population $0\leq p_j\leq 1$ transferred between the momentum states can be controlled via the interaction duration between the BEC and the laser beams \cite{Neumann2021}. To avoid unwanted transitions outside of the FWM states, the preparation sequence shown in Fig. \ref{fig:bragg_diffraction} and Table \ref{tab:preparation_sequence} is developed. For visualization, we choose $\vec{k}_1=\hat{\vec{k}}_x$, $\vec{k}_2=-\hat{\vec{k}}_x$, $\vec{k}_3=\hat{\vec{k}}_y$ and $\vec{k}_4=-\hat{\vec{k}}_y$. However, all combinations fulfilling (\ref{eq:4WM_cond}) can be prepared by the described procedure. After the pulse sequence, the total particle number is transferred into the FWM states,
\begin{align}
    (1-p_3)(1-p_2)+p_3(1-p_2)+(1-p_4)p_2+p_4p_2=1, \label{eq:number_conservation_preparation}
\end{align}
and all combinations can be realized (see Fig. \ref{fig:preparation_sequence}).

\bibliography{bib_4WM_neuron} 

\end{document}